\documentclass[letterpaper]{article} 

\usepackage{aaai23}  
\usepackage{times}  
\usepackage{helvet}  
\usepackage{courier}  
\usepackage[hyphens]{url}  
\usepackage{graphicx} 
\usepackage{natbib}  
\usepackage{caption} 
\usepackage{algorithm}
\usepackage{algorithmic}
\usepackage{amsmath, amssymb, amsfonts}
\usepackage[inline]{enumitem}
\usepackage{newfloat}
\usepackage{listings}
\DeclareCaptionStyle{ruled}{labelfont=normalfont,labelsep=colon,strut=off} 
\floatstyle{ruled}
\newfloat{listing}{tb}{lst}{}
\floatname{listing}{Listing}
\usepackage{xcolor}
\usepackage{amsmath}
\usepackage{amssymb}
\usepackage{amsthm}
\usepackage{cleveref}
\usepackage[subrefformat=parens]{subcaption}
\Crefname{ALC@unique}{Line}{Lines}

\usepackage{tikz}
\usetikzlibrary{automata, positioning, arrows, calc}
\tikzset{initial text={}} 
\usepackage{changes}
\definechangesauthor[color=orange]{KW}
\definechangesauthor[color=blue]{ML}
\definechangesauthor[color=purple]{SS}
\title{Timed Partial Order Inference Algorithm}
\author {
    Kandai Watanabe \textsuperscript{\rm 1}
    Bardh Hoxha \textsuperscript{\rm 2} 
    Danil Prokhorov \textsuperscript{\rm 2}
    Georgios Fainekos \textsuperscript{\rm 2} 
    Morteza Lahijanian \textsuperscript{\rm 1}
    Sriram Sankaranarayanan \textsuperscript{\rm 1}
    Tomoya Yamaguchi \textsuperscript{\rm2, \rm 3}
}
\affiliations {
    \textsuperscript{\rm 1} University of Colorado Boulder\\
    \textsuperscript{\rm 2} Toyota Research Institute North America\\
    \textsuperscript{\rm 3} Woven Planet\\
    first.lastname@colorado.edu,\
    first.lastname@toyota.com
}

\usepackage{xspace}

\newtheorem{theorem}{Theorem}

\newtheorem{definition}{Definition}

\newtheorem{example}{Example}

\newcommand{\events}{\Pi}
\newcommand{\event}{e}

\newcommand{\timedtrace}{\tau}

\newcommand{\PO}{P}

\newcommand{\po}{PO\xspace}
\newcommand{\tpo}{TPO\xspace}

\newcommand\reals{\mathbb{R}}
\begin{document}
\maketitle

\begin{abstract}

In this work, we propose the model of timed partial orders (TPOs) for specifying workflow schedules, especially for modeling manufacturing processes. TPOs integrate partial orders over events in a workflow, specifying ``happens-before'' relations, with timing constraints specified using guards and resets on clocks --- an idea borrowed from timed-automata specifications. TPOs naturally allow us to capture event ordering, along with a restricted but useful class of timing relationships. Next, we consider the problem of mining TPO schedules from workflow logs, which include events along with their time stamps. We demonstrate a relationship between formulating TPOs and the graph-coloring problem, and present an algorithm for learning TPOs with correctness guarantees.
We demonstrate our approach on synthetic datasets, including two datasets inspired by real-life applications of aircraft turnaround and gameplay videos of the Overcooked computer game. Our TPO mining algorithm can infer TPOs involving hundreds of events from thousands of data-points within a few seconds. We show that the resulting TPOs provide useful insights into the dependencies and timing constraints for workflows.
\end{abstract}

\section{Introduction}
Workflows appear in diverse areas, including business processes \cite{agrawal1998mining, datta1998automating}, software engineering \cite{cook1998discovering}, and factory pipelines \cite{choudhary2009data}. The individual events in a workflow, such as the start/end of a particular task or the achievement of an intermediate sub-goal, are ordered according to a strict partial order that specifies that some event $e_i$ always \emph{happens before} another event $e_j$. Such partial orders have been used to represent plans in classic AI planning algorithms~\cite{russell:2009:AI}.
Beyond partial orders, we often have timing constraints between events in a workflow that place bounds on the time when an individual event occurs or the time elapsed between two events in the workflow. In this paper,
we tackle the key problems of specifying these timing constraints in a succinct manner and mining such schedules from data that consists of sequences of time-stamped events, each representing an execution
of the workflow.  Such a specification enables us to implement workflow monitoring algorithms, understand sources of timing uncertainties in workflows and optimize the workflow in order to realize cost savings.

In this paper, we introduce a model called Timed Partial Order (\tpo), and propose a novel algorithm to infer a \tpo from timestamped event sequences. The \tpo  model integrates partial orders with timing constraints that are expressed with clocks, whose idea comes from timed automata specifications \cite{Alur+Dill/1994/Theory}. The clocks act as timers that express bounds on the time intervals between pairs of events. We first introduce the TPO model and present an analysis of its expressivity. We show that TPOs can succinctly capture a complex set of timing constraints by checking assertions over clocks and selectively resetting these clocks when certain events happen. In particular, we identify structural constraints that yield a restricted class called \emph{race-free} TPOs. We show that race-free TPOs correspond precisely to difference constraints involving time intervals between events. We demonstrate an algorithm that translates a system of explicit constraints on the timing between events into a race-free TPO specification. Next, we solve the problem of mining race-free TPOs from timestamped event sequences. Our approach first mines  partial order specification and timing constraints from the data, translating these constraints into race-free TPO specifications.

We present an evaluation of our approach on a combination of synthetic benchmarks to show that our algorithm can process large numbers of event sequences in a matter of seconds, and provides succinct TPO specifications in terms of the number of clocks needed. Next, we demonstrate our approach on two examples inspired by real-life applications: a model of events involved in the workflow for commercial aircraft turnaround and an analysis of the multiplayer computer game Overcooked, as played by beginner and expert players. In both of these examples, we use TPO mining to produce specifications that can yield useful insights about the nature of the workflows in question.

\begin{figure}
\begin{center}
\begin{tikzpicture}
\begin{scope}
\matrix[every node/.style={draw=black, rectangle, rounded corners}, row sep=5pt, column sep=7pt]{
                & \node(n2){$e_2$}; && \node(n3) {$e_3$}; \\
 \node(n1){$e_1$}; &  & & & \node(n5){$e_5$}; & \node(n6){$e_6$};\\
 & & \node(n4){$e_4$}; & \\
};
\path[->, line width=2pt] (n1) edge (n2)
(n2) edge (n3)
(n3) edge (n5)
(n4) edge (n5)
(n3) edge (n5)
(n5) edge (n6)
(n1) edge (n4);
\end{scope}
\begin{scope}[xshift=4.5cm]
\node at (0,0) {
\begin{tabular}{ll}
\hline
$e_1$ & Car Arrived \\
$e_2$ &  Add Glue Primer \\
$e_3$ & Grip/Lift Window \\
$e_4$ & Clean Dust \\
$e_5$ & Place Window \\
$e_6$ & Release Grippers\\
\hline
\end{tabular}
};
\end{scope}
\end{tikzpicture}
\end{center}

 \caption{Partial Order for a windshield installation task in an automobile manufacturing facility. Events $e_1, \ldots, e_5$ represent events such as ``car arrived'' ($e_1$) or the commencement of various tasks such as ``clean dust'' ($e_4$).}
    \label{fig:RunningExample}
\end{figure}

\begin{example}\label{ex:windshield-example}
Figure~\ref{fig:RunningExample} shows the set of events that define the (simplified) process of placing a windshield on a car in an automobile manufacturing facility. The events are described along with a directed graph that represents the partial order between events. For instance, the edge from $e_2$ to $e_3$ specifies that the event $e_2$ must always precede $e_3$ for a (successful) windshield installation. However, like many tasks in an assembly line, there are timing constraints that must be respected. Some of the timing constraints  for the windshield installation task are summarized in the table below:

\begin{center}
\begin{tabular}{llll}
\hline
ID & Interval & Cons. & Remark \\
\hline
$C_1:$ & $e_2 \rightarrow e_5$ & $\leq 40s$ & Glue appl. to place window\\
$C_2:$ &$e_5 \rightarrow e_6$ & $\geq 30s$ & Min. glue setting time \\
$C_3:$ &$e_1 \rightarrow e_4 $ & $\leq 5s$ & Max.  time to start cleaning \\
$C_4:$ &$e_1 \rightarrow  e_6$ & $ \leq 100s$ & Max. end-to-end time.\\

\hline
\end{tabular}
\end{center}
$C_1$ enforces that once the glue primer is applied to the window ($e_2$), the window must be placed on the car ($e_5$) within $40$ seconds. Similarly, $C_2$  states that once the window is placed on the car, a wait of at least $30$ seconds is required for the glue to set before releasing the grippers.
\end{example}

In this paper, we first describe the model of \emph{timed partial orders}, which combines partial orders between events with ideas from timed automata to represent timing constraints. Next, we show how, given a dataset of time-stamped events, we can mine the timed partial order specification.

\section{Related Work}

Learning a task from log data is known as \textit{Workflow Mining} or \textit{Process Mining}. Most approaches to process mining focus on modeling the order in which tasks are performed, but do not capture timing constraints. As such, process mining has been widely used in the industry \cite{huser2012process, van2013process}. It is supported by commercial tools such as Disco, Celonis, and Process Gold; and  open source tools  such as ProM, Apromore and pm4py \cite{berti2019process}. Under the hood, these tools implement algorithms such as the $\alpha$-algorithm that outputs a Petri Net \cite{petri1962kommunikation}, $\alpha\!+$  algorithm that can handle loops \cite{van2004workflow}, and an algorithm that can handle duplicate events \cite{herbst2000machine}.
Various types of $\alpha$-algorithm were introduced \cite{van2009process} to overcome some of its limitations.
Other approaches are also introduced such as region-based approaches \cite{van2010process, carmona2008region} that can express more complex control-flow structures and heuristic mining \cite{weijters2003rediscovering}, fuzzy mining \cite{gunther2007fuzzy}, query-based mining \cite{esparza2010learning} that can handle incomplete data and genetic process mining \cite{de2007genetic} that can handle noise.
The problem of learning timing constraints has been studied, as well. \citet{berlingerio2009temporal} focuses on inferring ``typical transition time'' between two events by counting the number of steps between them.  \citet{sciavicco2021mining} mine \emph{Conditional Simple Temporal Network with Uncertainty and Decisions}  (CSTNUD) models from log data. CSTNUD are a temporal networks (Cf.\cite{dechter1991temporal}), wherein timing differences between events are represented as ``durations''. These durations  can also be viewed as a single clock that resets at every transition.  This representation is frequently used in works such as \citet{verwer2012efficiently}. However, their expression is limited to a single clock, whereas our approach uses multiple clocks that need not be reset at each transition. Multiple clocks are necessary in order to capture more complex timing constraints that are frequently seen in our examples and case-studies.
Moreover, their method assumes that dependencies between events are manually provided in the log, e.g., event $E$ happens 5 seconds after $A$, and only depends on \emph{one} event. In contrast, our method can mine the structure at the same time and allows events to depend on multiple events. Moreover, we enforce timing constraints with ``clocks'' for easier interpretations and faster computation when planning.


Automata such as timed-automata ~\cite{Alur+Dill/1994/Theory} can be used to model
workflow schedules.
Researchers have extended automata learning techniques to learn timed automata that can capture timing constraints.  For instance, \citet{verwer2012efficiently} extended the Evidence-Driven State Merging (EDSM)-based algorithms originally proposed by \citet{gold1978complexity} to learn from data with timestamps and estimate a real-time timed automaton whose edges are labeled with time duration. Although the algorithm is fast, it can only infer simple time constraints. \citet{an2020learning} also extended the L$^*$-based approach of \citet{angluin1987learning}, and \citet{tappler2022timed} formulated the problem such that it can be solved by Satisfiability Modulo Theories (SMT). However, the former assumes a perfect oracle, and the latter is often very slow due to the nature of the SMT solver, which often leads to exponential time. The genetic algorithm-based approach \cite{tappler2019time} is fast and gives a good solution if it finds one, but has no optimality guarantees. All these approaches can estimate some types of Timed Automata, which are very expressive models, but are disadvantageous due to the fundamental difficulty of learning timed automata from trace data.

\section{Timed Partial Orders}

In this section, we first define the preliminary concepts of a timed partial order and show how it can specify timing constraints between events in a manufacturing workflow.

We model workflows as a timed sequence involving a fixed number of \emph{events}. These events may include the start/finish of a given task, or the achievement of a certain physical condition, e.g., the temperature of the water has exceeded 100$^\circ$C. We assume that the set of events are fixed \emph{a priori}. Furthermore, we assume that  repetitions of events are \emph{disambiguated} by giving them unique labels.

Let $\events = \{\event_1, \ldots, \event_n \}$ be the set of events.  A given \emph{run} of the workflow is defined by  a \emph{timed trace} $\timedtrace: \events \rightarrow \reals_{\geq 0}$  that maps each event with a non-negative timestamp, i.e.,
\[ \timedtrace= \{ \event_1 \mapsto t_1, \event_2 \mapsto t_2, \ldots, \event_n \mapsto  t_n \}  \,,\]
wherein  $t_i \geq 0$ denotes the timestamp for event $\event_i \in \events$.

A timed-trace $\timedtrace$ induces an ordering over the events according to increasing time stamps, and thus may also be viewed as a sequence:
$(\sigma_1, t^{(1)}), \ldots, (\sigma_n, t^{(n)})$,
wherein each $\sigma_i \in \events$ denotes a unique $i^{th}$ event in the trace with corresponding time stamp $t^{(i)}$ and furthermore, $t^{(1)} < t^{(2)} < \cdots < t^{(n)}$. 
Below, we formulate models for the possible timed traces corresponding to runs of a workflow.

A \textit{(strict) Partial Order} (\po) $\PO$ is a relation $\prec$ on a set $\events$ that is irreflexive, asymmetric, and transitive. We write $e_i \preceq e_j$ if $e_i \prec e_j$ or $i = j$. If $\event_i \prec \event_j$ holds, then for any timed trace $\timedtrace$ we will require that $\tau(\event_i) \leq \tau(\event_j)$.
We now describe the model of timed POs, which combine POs that specify happens-before relations between events with timing constraints.

\begin{definition}[Timed Partial Order]
A timed partial order (TPO) is specified by a directed-acyclic graph (DAG) $G: (\events, \prec)$ describing a strict partial order over $\events$ augmented with the following:
\begin{enumerate}
    \item A finite set of \emph{clocks} $C = \{ c_1, \ldots, c_m \}$,
    \item A \emph{guard map} $G$ that maps each event $\event_i$ to a guard condition, which is a conjunction of the form
$G(\event_i): \bigwedge_{j=1}^{n_i} c_j \bowtie a_j$ wherein $c_j \in C$ denotes a clock, $\bowtie \in \{ \leq, \geq \}$, and $a_j \in \reals_{\geq 0}$ is a non-negative constant,
\item A \emph{reset map} $R: \events \to 2^{C}$ that associates each event $\event_i$ with a subset of clocks $R(\event_i) \subseteq C$ that are to be reset to $0$ whenever event $\event_i$ is encountered.
\end{enumerate}
\end{definition}

A valuation $\nu: C \rightarrow \reals_{\geq 0}$ assigns each clock $c_i \in C$ to a non-negative number $\nu(c_i)$. A given valuation $\nu$ can be advanced in time by a fixed  $\delta \geq 0$ to yield a new valuation  $\nu'\ :=\ \nu \oplus \delta$ such that $\nu'(c_j) = \nu(c_j) + \delta$ for all $c_j \in C$. Likewise, given a valuation $\nu$ and a subset of clocks $\hat{C} \subseteq C$, we denote the valuation $\nu' := \textsf{reset}(\nu, \hat{C})$ as that obtained by setting each clock $c \in \hat{C} $ to be $0$:
$\nu'(c) = \begin{cases}
0 & c \in \hat{C} \\
\nu(c) & c \not \in \hat{C} \end{cases}$.
Let $\nu_0$ represent a fixed special initial valuation wherein $\nu_0(c_j) = 0$ for all clocks and let $t^{(0)} = 0$. Timestamps represent global time since the inception of the process whereas clocks measure the time elapsed since their last reset.

\begin{definition}[Semantics of Timed Partial Orders]\label{def:tpo-semantics}
A \emph{run} of a timed-partial order is a sequence of triples
\[ \rho: (\sigma_1, t^{(1)}, \nu_1), \ldots, (\sigma_n, t^{(n)}, \nu_n) \,,\]
wherein, each $\sigma_i \in \events$, $\sigma_i \not= \sigma_j$ for $i \not= j$, and \item each $\nu_j$ is a valuation of clocks $C$.
\begin{enumerate}
    \item Time stamps are non-decreasing: $t^{(1)} \leq t^{(2)} \leq \cdots \leq t^{(n)}$. Let $\Delta t^{(j)}$ denote the difference $t^{(j)} - t^{(j-1)}$ for $j \in \{ 1, \ldots, n \}$ (note that $t^{(0)} = 0$).
    \item The sequence $\sigma_1, \ldots, \sigma_n$ is a linearization of the partial order $\prec$: if $\sigma_a \prec \sigma_b$ holds then $a < b$.

\item For each $j \in \{1, \ldots, n \}$,  the valuation given by $\nu_{j-1} \oplus \Delta t^{(j)} $ satisfies the guard condition $G(\sigma_j)$.
\item The valuation $\nu_j$ must equal $\textsf{reset}(\nu_{j-1} + \Delta t^{(j)}, R(\sigma_j))$. I.e,  we allow time $\Delta t^{(j)}$ to elapse and then reset the clocks in $R(\sigma_j)$.
\end{enumerate}

\end{definition}

A timed trace $\timedtrace$ viewed as a sequence $(\sigma_1, t^{(1)}), \ldots, (\sigma_n, t^{(n)})$ is \emph{compatible} with a timed partial order specification iff there exists a run  of the form
$(\sigma_1, t^{(1)}, \nu_1), \ldots, (\sigma_n, t^{(n)}, \nu_n)$.

\begin{figure}
\begin{tikzpicture}
\begin{scope}
\matrix[every node/.style={draw=black, rectangle, rounded corners}, row sep=5pt, column sep=15pt]{
                & \node(n2){$e_2$}; && \node(n3) {$e_3$}; \\
 \node(n1){$e_1$}; &  & & & &\node(n5){$e_5$}; &  & \node(n6){$e_6$};\\
 & & \node(n4){$e_4$}; & \\
};
\path[->, line width=2pt] (n1) edge (n2)
(n2) edge (n3)
(n3) edge (n5)
(n4) edge (n5)
(n3) edge (n5)
(n5) edge (n6)
(n1) edge (n4);
\draw (n1.north)++(0,0.3) node{ \footnotesize $c_1 := 0$};
\draw (n4.north)++(0,0.3) node{ \footnotesize $c_1 \leq 5$};
\draw (n6.north)++(0,0.6) node{ \footnotesize $\begin{array}{c}
c_1 \leq 100\ \land\\
c_2 \geq 30\end{array}$};
\draw (n2.north)++(0,0.3) node{ \footnotesize $c_2 := 0$};
\draw (n5.north)++(0,0.6) node{ \footnotesize $\begin{array}{c} c_2 \leq 40 \\
\rightarrow\ c_2 := 0 \\ \end{array}$};

\end{scope}
\end{tikzpicture}
\caption{TPO for the car windshield installation workflow.}\label{fig:tpo-running-example}
\end{figure}
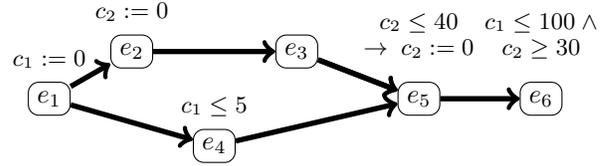

\begin{example}\label{ex:windshield-tpo}
Figure~\ref{fig:tpo-running-example} shows the TPO specification for the
windshield installation task in Example~\ref{ex:windshield-example}.
Two clocks $c_1, c_2$ are used to enforce the constraints $C_1, \ldots, C_4$ from that example. First, we use clock $c_1$ which is  reset to zero at the very beginning when event $e_1$ occurs. The guard $c_1 \leq 5$ at event $e_4$ checks that the timing interval between $e_1$ and $e_4$ is at most $5$ time units ($C_3$). Likewise, the guard $c_1 \leq 100$ at event $e_6$ ensures constraint $C_4$. The clock $c_2$ is reset first when event $e_2$ occurs. The guard $c_2 \leq 40$ associated with $e_5$ enforces constraint $C_1$. The clock is then reset to zero as part of the same event and then further the guard $c_2 \geq 30$ associated with $e_6$ ensures constraint $C_2$.
\end{example}

\subsection{Expressivity of TPOs}
We will now examine the expressivity of TPOs. In general,  TPOs enforce timing constraints using clocks. We derive key insights into the nature of these timing constraints. Furthermore, we define useful structural restrictions to TPOs that make the problem of reasoning about their behavior easier. This will pave the way for the TPO mining algorithm that will be presented in the subsequent section.

\begin{example}[Limits on Expressivity]
Suppose, for the windshield installation task in Example~\ref{ex:windshield-example}, we wish to add a constraint ---
{the time taken to clean the windshield (time elapsed between events $e_4$ and $e_5$) must be less than the time taken to add the glue primer to the window (time elapsed between events $e_2$ and $e_3$)}.  Such a  constraint compares the intervals between two sets of events. As such, this will not be expressible in the formalism of TPOs. The reason (in part) is that we disallow the guards to compare the values of clocks.
\end{example}

For the remainder of this section, let us fix a TPO with events $\Pi = \{ e_1, \ldots, e_n \}$ and partial order relation
$\prec$,  clocks $C = \{ c_1, \ldots, c_m \} $, guard map $G$ and resets $R$. We will represent a ``generic'' timed trace $\timedtrace: \{ e_1 \mapsto t_1, \ldots, e_n \mapsto t_n \}$ as a vector  $(t_1, \ldots, t_n) \in \reals^n$, wherein $t_i$ denotes the time at which the event $e_i$ occurs.

\begin{definition}[Constraints representing a TPO]
An assertion $\varphi[t_1, \ldots, t_n]$ involving $t_1, \ldots, t_n$ represents a
TPO iff  every timed trace $\timedtrace: \{ e_1 \mapsto t_1, \ldots, e_n \mapsto t_n \}$  compatible with the TPO satisfies $\varphi$, and conversely, every timed trace satisfying $\varphi$ is compatible with the TPO.
\end{definition}

For the partial order $\prec$, let $\psi_{POR}$ denote
the assertion
\begin{equation}\label{eq:psi-por}
   \psi_{POR}:\  \bigwedge_{e_i \prec e_j} t_j - t_i \geq 0\ \land\ \bigwedge_{j=1}^n t_j \geq 0
\end{equation}
 expressing the timing constraints of $\prec$.

\begin{example}
The TPO in Ex.~\ref{ex:windshield-tpo} is represented by constraints:
\[ \footnotesize \left( \begin{array}{c}
t_1 \leq t_2 \leq t_3 \leq t_5 \ \land \
t_1 \leq t_4 \leq t_5 \leq t_6 \ \land\\
 t_5 - t_2 \leq 40\ \land\ t_6 - t_5 \geq 30\ \land t_4 - t_1 \leq 5\ \land\ t_6 - t_1 \leq 100
 \end{array} \right)  \]
 The first row represents the relation $\prec$ from the partial order whereas the last  row represents the timing constraints $C_1, \ldots, C_4$ discussed in Example~\ref{ex:windshield-example}. Note that (a) TPOs provide a more succinct representation of the timing constraints; and (b) they also specify a procedure to monitor the timed-trace by maintaining some clocks, checking guards on them upon events and resetting the clocks as specified by the TPO. Timing constraints on the other hand require us to potentially store the times of each and  every event in the trace.
\end{example}
Let us consider another example below.
\begin{example}\label{Ex:racy-tpo-example}
Consider a TPO over $3$ events $e_1, e_2, e_3$ with the $\prec = \emptyset$. In other words, there is no fixed ``happens-before'' order between these events.
\begin{center}
\begin{tikzpicture}
\matrix[every node/.style={draw=black, rectangle, rounded corners}, row sep=5pt, column sep=68pt]{
\node(n1){$e_1$}; & \node(n2){$e_2$}; & \node(n3){$e_3$};\\
};
\draw (n1.north)++(0,0.2) node {\footnotesize $\begin{array}{l}
c_1 \leq 1 \rightarrow\
c_1 := 0 \end{array}$};
\draw (n2.north)++(0,0.2) node {\footnotesize $\begin{array}{l} c_1 \leq 1\ \rightarrow\  c_1 := 0\end{array}$};
\draw (n3.north)++(0,0.2) node {\footnotesize $\begin{array}{l} c_1 \leq 1\ \rightarrow c_1 := 0\end{array}$};
\end{tikzpicture}
\end{center}
The TPO has a single clock $c_1$, with each event having a guard $c_1 \leq 1$ and resetting the clock $c_1$ to $0$. The set of admissible timed traces ensure that: (a) the events $e_1, \ldots, e_3$ may happen in arbitrary order; (b) the first event must appear within $1$ time unit of the start of the process; (c) each subsequent event must happen within $1$ time unit of the previous event. The following constraints represent this TPO:
\[ \begin{array}{l}
(t_1 \leq t_2 \leq t_3\ \Rightarrow\ t_1 \leq 1\ \land\ t_2 - t_1 \leq 1\ \land\ t_3 - t_2 \leq 1)\land \\
(t_1 \leq t_3 \leq t_2\ \Rightarrow\ t_1 \leq 1\ \land\ t_3 - t_1 \leq 1\ \land\ t_2 - t_3 \leq 1)\land \\
\cdots \\
(t_3 \leq t_2 \leq t_1\ \Rightarrow\ t_3 \leq 1\ \land\ t_2 - t_3 \leq 1\ \land\ t_1 - t_2 \leq 1) \\
\end{array}\]
Each of the 3! clauses correspond to a linearization of the partial order which leads to different constraints between intervals over successive events as dictated by the TPO.
\end{example}

To avoid the need to reason over linearization of the underlying partial order, we define a structurally restricted class of TPOs that we call ``race-free TPOs''. The race-freedom here denotes that the timing constraints remain independent of the actual order in which the events occur. We say that an event $e_i$ in a TPO is \emph{dependent} on a clock $c_j$ iff the guard for $e_i$ refers to  $c_j$ or  $c_j$ is reset by $e_i$.

\begin{definition}[Race-Free TPOs]
A TPO is said to be race-free iff for every clock $c \in C$ and two different events  $e_i, e_j$ that are dependent on $c$, either $e_i \prec e_j$ or $e_j \prec e_i$. In other words, events that are independent according to the partial order (parallel events) do not refer to the same clock.
\end{definition}

Note that the TPO in Example~\ref{ex:windshield-tpo} is clearly race-free.
For instance, events $e_2, e_5$ refer to clock $c_2$ but $e_2 \prec e_5$. This can be checked for every clock and every pair of events that are dependent on that clock. However, the TPO in Example~\ref{Ex:racy-tpo-example} is not race-free. Events $e_1, e_2, e_3$ all refer to clock $c_1$ but we do not have any precedence relationship between any of them.

We will now establish the key theorem that characterizes the timing constraints corresponding to race-free TPOs.
\begin{theorem}\label{Theorem:tpo-expressivity-race-free}
A (race-free) $\tpo$ can be represented by a conjunction of inequalities of the form:
\[ \bigwedge_{i, j\ \mathsf{s.t.}\ e_i \prec e_j}\ (t_j - t_i) \in [\ell_{j,i}, u_{j,i} ] \ \land\ \bigwedge_{j=1}^n t_j \in [a_j, b_j ] \,,\]
wherein $\ell_{i,j} \geq 0, a_j \geq 0$ form lower bounds and $u_{j,i}, b_j \in \reals_{\geq 0} \cup \{ \infty \} $ are upper bounds that can be non-negative real numbers as well as $ +\infty$.
\end{theorem}
Full proofs are provided in the appendix. Briefly, the theorem holds because in a race-free TPO, a clock $c$ that is reset at some event $e_i$ and subsequently referred to at event $e_k$ (without intervening reset) ensures that $e_i \prec e_k$. Therefore, the clock $c$ at any point refers to the time difference $t_k - t_i$. Therefore, guards on clocks translate into constraints involving differences $t_k - t_i$. However, if a clock is not reset, it measures the time elapsed since the start of the process. A guard on such a clock is simply a constraint on $t_k$.

\subsubsection{Constructing Timed Partial Orders From Constraints}

The main insight behind our paper lies in proving the converse of Theorem~\ref{Theorem:tpo-expressivity-race-free}. Let us fix a set of events $\Pi = \{ e_1, \ldots, e_n\}$ and a partial order $\prec$ between them.
\begin{theorem}
Given timing constraints of the form
\begin{equation} \label{eq:timing-constraints}
\varphi:\ \bigwedge_{j=1}^n t_j \in [a_{j}, b_j]\ \land\ \bigwedge_{e_i \prec e_j} (t_{j} - t_i) \in [\ell_{i,j}, u_{i,j} ] \,,
\end{equation}
wherein $a_j, \ell_{i,j} \geq 0$ and $b_j, u_{i,j} \in \reals_{\geq 0} \cup \{ \infty \}$, there is a race-free TPO that represents the timing constraints $\varphi$.
\end{theorem}
The proof of this theorem lies in the procedure we will now present  to synthesize such a TPO  involving three major steps:
(a) Remove \emph{redundant} timing constraints from $\varphi$ to obtain an irredundant representation $\tilde{\varphi}$; (b) Construct a clock allocation graph $G$ from
$\tilde{\varphi}$; (c) solve a graph coloring problem on $G$ and (d) translate the
graph coloring result into clocks, clock guards and resets for the TPO.

\begin{example}\label{Ex:running-example-timing-constraints}
Consider once again the windshield installation task from Example~\ref{ex:windshield-example}.
Ignoring  $C_1, \ldots, C_4$, let us instead consider the following timing  constraints:
\begin{equation}\label{eq:example-timing-constraints}
\footnotesize
\varphi:\ \left(
\begin{array}{c}
t_3 - t_1 \in [10, \infty]\ \land\ t_5 - t_1 \in [0, 15]\ \land\\
t_5 - t_3 \in [0, 5] \ \land\ t_6 - t_5 \in [0, 8]\ \land \\
t_5 - t_4 \in [5, \infty]\ \land\ t_6 - t_4 \in [4, 10]  \\
\end{array}\right)
\end{equation}
\end{example}

\paragraph{Redundancy Elimination and Simplification:}  First, we introduce a fictitious initial event $e_0$ which always happens at fixed time $t_0 = 0$ such that $e_0 \prec e_j$ for all $j \in \{ 1, \ldots, n \}$. The constraints in Eq.~\eqref{eq:timing-constraints} are now written as:
\begin{equation}
\bigwedge_{e_i \prec e_j} \ell_{i,j} \leq (t_j - t_i) \land (t_j - t_i) \leq u_{i,j} \,.
 \end{equation}
Next, we eliminate redundant constraints of two types: (a) any constraint of the form $t_j - t_i \bowtie a$ that is implied by the conjunction of the remaining constraints; (b) trivial constraints  with $\ell_{i,j} = 0$ or  $u_{i,j} = \infty$;.
The result of redundancy elimination may be witten as:
\begin{equation}\label{eq:irredundant-constraint}
\tilde{\varphi}:\ \bigwedge_{i,j}\ \bigwedge_{k} (t_j - t_i) \bowtie l_{i,j,k},\ \mbox{wherein}\ \bowtie\ \in\ \{ \leq, \geq \} \,.
\end{equation}
where $k$ iterates over all inequalities that involve $t_j- t_i$. Additionally, for each inequality $t_j - t_i \bowtie l_{i,j,k}$, the corresponding events must satisfy $e_i \prec e_j$. We provide further details on redundancy elimination in the subsequent section.

\begin{example}\label{Ex:running-example-timing-constraints-2}
Consider the constraint~\eqref{eq:example-timing-constraints} in
Example~\ref{Ex:running-example-timing-constraints}.
The constraint $t_5 - t_3 \leq 5$ is redundant since we can infer it
from the two constraints $t_3 - t_1 \geq 10$ and $t_5 - t_1 \leq 15$.
Removing the trival and redundant constraints yields
\begin{equation}\label{eq:example-timing-reduced}
\tilde{\varphi}:\ \left(\begin{array}{c}
t_3 - t_1 \geq 10 \ \land\ t_5 - t_1 \leq 15 \ \land\ t_6 - t_5 \leq 8\\
t_5 - t_4 \geq 5 \ \land\ t_6 - t_4 \in [4,10] \\
\end{array}\right)\,.
\end{equation}
\end{example}

\paragraph{Allocating Clocks to Enforce Constraints:} In order to enforce a constraint of the form $t_j - t_i \bowtie l_{i,j,k}$ using clocks:
\begin{enumerate*}
\item Reset a ``dedicated'' clock $c_i$ at the same instant when event $e_i$ occurs;
\item Add the guard $c_i \bowtie l_{i,j,k}$ for event $e_j$.
\end{enumerate*}
In effect, $c_i$ measures time elapsed since event $e_i$. When event $e_j$ happens, its value equals $t_j - t_i$.
In fact, the clock $c_i$ can be used to enforce multiple conjunctions of the form $t_{j_1} - t_i \bowtie a_1\ \land\ \cdots \ \land\ t_{j_l} - t_i \bowtie a_l$ since
the structure of $\tilde{\varphi}$ (Eq.~\eqref{eq:irredundant-constraint}) guarantees that
$e_i \prec e_{j_1}, \ldots, e_i \prec e_{j_l} $. Thus, the modified strategy is as follows:
\begin{enumerate}[nosep]
\item Write $\tilde{\varphi}$ as $\tilde{\varphi}_0 \ \land\ \cdots\ \tilde{\varphi}_n$, wherein
$\tilde{\varphi}_i$ collects all inequalities in $\tilde{\varphi}$ of the form $(t_k - t_i) \bowtie a_{k,i}$.
\item If $\tilde{\varphi}_i$ is not empty, then allocate a dedicated clock $c_i$ that is reset at event $e_i$. In special case, since event $e_0$ is fictitious, we allocate the clock $c_0$ but do not reset it.
\item For the inequality  $(t_k - t_i) \bowtie a_{k,i}$ in $\tilde{\varphi}$, add the conjunction $c_i \bowtie a_{k,i}$ to the guard $G(e_k)$ for event $e_k$.
\end{enumerate}

Thus, the scheme so far constructs a TPO with at most $n+1$ clocks. Since $n$ can be quite large ($\sim 500$) for some manufacturing workflows, we  wish to minimize the number of clocks to reduce the complexity of the overall TPO.

\begin{example}\label{Ex:running-example-timing-constraints-3}
Continuing from
Example~\ref{Ex:running-example-timing-constraints-2}.
Following the technique presented thus far, we split the constraint $\tilde{\varphi}$
(Cf.~\eqref{eq:example-timing-reduced}) into three parts (underlining is for emphasis)
given by $\tilde{\varphi}_1 = \ (t_3 - \underline{t_1} \geq 10\ \land\ t_5 - \underline{t_1} \leq 15)$,
$\tilde{\varphi}_4 =  (t_5 - \underline{t_4} \geq 5 \ \land\ t_6 - \underline{t_4} \geq 4\ \land\ t_6 - \underline{t_4} \leq 10)$ and $\tilde{\varphi}_5=  ( t_6 - \underline{t_5} \leq 8)$. We allocate three clocks $c_1, c_4, c_5$ to track these three sets of constraints, respectively. Clock $c_1$ is reset at event $e_1$,  $c_4$ at event $e_4$ and $c_5$  at event $e_5$.  The following guards are added:
\[{\footnotesize \begin{array}{ll}
\text{Constraint} & \text{Guard}\\
\hline
t_3 - t_1 \geq 10\ \land\ t_5 - t_1 \leq 15 & c_1 \geq 10 @ e_3,\ c_1 \leq 15 @ e_5 \\
t_5 - t_4 \geq 5 \ \land\ t_6 - t_4 \in [4,10] & c_4 \geq 5 @ e_5, (c_4 \in [4,10]) @ e_6 \\
t_6- t_5 \leq 8 & c_5 \leq 8 @ e_6 \\
\hline
\end{array}}\]
\end{example}

\paragraph{Minimizing Clocks in the TPO:} In order to reduce the number of clocks, we ask the following question: under what conditions can we ``reuse'' the clock $c_i$ corresponding to event $e_i$ for a different event $e_j$?

For any given event $e_i$, let $e_k$ be the event that is maximal according to the precedence relation $\prec$ such that a timing constraint of the form $(t_k - t_i) \bowtie a_{k,i}$ exists in $\tilde{\varphi}$. If no such inequality exists in the first place, then clock $c_i$ would not exist in the first place. We can reuse the clock $c_i$ for any ``later'' event $e_j$ such that  $e_k \preceq e_j$, since the last time clock $c_i$ is used is at event $e_k$. Recall, from  TPO semantics  in Def.~\ref{def:tpo-semantics}, that clocks are  reset only  \emph{after} the guards are checked.

We construct a \emph{clock allocation graph} $G_c$, an undirected graph
whose vertices are the clocks considered so far.
\begin{enumerate}[nosep]
    \item Corresponding each clock $c_i$, we compute its latest guarded event $L(i)$ as follows:
     \begin{enumerate}[nosep]
     \item Let $E_i =  \{ e_k\ |\ \text{inequality }\ t_k - t_i \bowtie a\ \text{in}\ \tilde{\varphi} \}$.
     \item Set $\text{L}(i) = \mathsf{sup}_{\prec} (E_i) $, the supremum in $E_i$ according to the $\prec$ order. We observe that clock $c_i$ can be reused after event $L(i)$ has occurred.
    \end{enumerate}
 \item Add an undirected edge $(c_i, c_j)$ whenever $L(i) \not\preceq e_j$ and $L(j) \not\preceq e_i$ enforcing that $c_i$ and $c_j$ be kept distinct.
\end{enumerate}

Recall that the graph coloring problem seeks to assign one of $m$ colors to each vertex of an
undirected graph so that no two vertices connected by an edge have the same color.
The main idea behind minimizing clock usage is to examine the optimal coloring of the
graph and whenever two nodes $c_i, c_j$ are the same color, we can substitute the use of clock $c_j$ by $c_i$. This process ensures that we use as many clocks as the number of colors used in graph coloring.
\begin{theorem}
If the clock allocation graph $G_c$ can be colored using $m$ colors, then we can construct a TPO with at most $m$ clocks to represent the timing constraints in $\tilde{\varphi}$.
\end{theorem}

\begin{example}
Continuing with Example~\ref{Ex:running-example-timing-constraints-3}, we compute the clock allocation graph $G_c$ with vertices $\{c_1, c_4, c_5\}$. Note that $L(1) = e_5$,
$L(4) = e_6$, and $L(5) = e_6$ as defined above.  Thus, according to the construction above, the clock allocation graph has two edges $\{ (c_1, c_4), (c_4, c_5) \}$.  This can be colored with  two colors and in particular nodes $c_1$ and $c_5$ have the same color. This denotes that the clock $c_5$ can be replaced with $c_1$ everywhere.
\end{example}

Note that the problem of checking if a graph $G$ may be colored using $m \geq 3$ colors is known to be NP-complete~\cite{Garey+Johnson/1979/Computers}. Nevertheless, graph coloring has been studied for numerous applications including notably register allocation for compilers and  scheduling problems~\cite{Chaitin+Others/1981/Register,Lotfi+Others/1986/Graph}. We may employ a simple greedy algorithm for graph coloring that guarantees that the number of colors is bounded by $1 + \Delta(G_c)$, wherein $\Delta(G_c)$ denotes the maximum number of neighbours for any vertex in $G_c$.

\subsection{Mining TPO from Timed Traces}

Given timed traces $\timedtrace_1, \ldots, \timedtrace_n$ by observing
some workflow with a fixed set of events $\Pi = \{ e_1, \ldots, e_n \}$, we wish to synthesize a TPO  such that all the timed traces in the given data $D$ are compatible with the TPO. To do so, requires identifying the partial order $\prec$, the clocks, guards and resets.

Our proposed approach proceeds in three steps: (a) identify the partial order information from the  timed traces;  (b) compute the tightest possible timing constraints of the form ~\eqref{eq:timing-constraints} that includes all the data; and (c) mine a TPO from the timing constraints in step (b) using the algorithm described in the previous section. Note that steps (a) and (b) are based on well-known techniques that will be briefly recalled in this section. We briefly describe these steps, concluding with a description of our implementation.

\paragraph{Partial Order Identification}
 In order to identify the partial order $\prec$,  we set $e_i \prec e_j$ for events $e_i, e_j \in \Pi$ iff in all the timed traces the event $e_i$ precedes $e_j$.

\paragraph{Formulating Timing Constraints} In order to formulate timing constraints,
we translate each timed trace in the data into a vector $(t_1, \ldots, t_n)$. Next, we consider bounds of two types: (a) Upper/lower bounds on each event time $t_i$ by itself to yield intervals $t_i \in [a_i, b_i]$, and (b) Upper/lower bounds on the time difference
$t_j - t_i$ whenever $e_i \prec e_j$ holds.

Whereas the interval bounds are simply the maximum and minimum values in the data, there are two drawbacks: (a) The process assumes that all upper bounds are finite since we can never infer a constraint of the form $t_j - t_i \in [a, \infty)$ from the data. However, there are statistical tests from \emph{extreme value theory} that can identify whether a distribution has an infinite support~\cite{Haan/2006/Extreme}. The application of these techniques relies on having a large volume of data. (b) The bounds themselves depend intimately on the amount of data and the sampling method. To mitigate this, we refer the reader to ideas from conformal prediction that allow us to bloat the intervals obtained from data appropriately to achieve a prediction with some associated confidence~\cite{Balasubramanian/2014/Conformal}.

\paragraph{Redundancy Elimination}
To eliminate redundancies, we formulate a series of optimization problems involving the constraints in \Cref{eq:timing-constraints}. We iterate through each inequality $(t_j - t_i) \leq  u_{i,j}$ (alternatively, $t_j - t_i \geq l_{i,j}$ ) from the system  in some order and carry out the following steps:
\begin{enumerate}[nosep]
    \item Remove the selected inequality and set the objective to maximize (alternatively, minimize) $t_j - t_i$ subject to the remaining constraints.
    \item If the resulting optimal value is $>  u_{i,j}$ (alternatively, $< l_{i,j}$) then the constraint is irredundant, and needs to be added back to the problem.
    \item Otherwise, the constraint is removed once and for all.
\end{enumerate}
The optimization problem in question is a linear programming problem that can be solved quite efficiently for the
special class of difference constraints encountered here.
However, the result of redundancy elimination varies, depending on the order of constraints in which we
process the inequalities. For example, the constraint $t_5 - t_1 \leq 15$ in \Cref{Ex:running-example-timing-constraints-2} can be removed instead of  $t_5 - t_3 \leq 5$. The problem of finding the set of irredundant constraints of the least cardinality is known to be NP-hard following a reduction from the minimum equivalent graph problem~\cite{Garey+Johnson/1979/Computers}. Therefore, we consider various
heuristics for deciding the order in which the constraints are to be considered.
\begin{enumerate}[nosep]
\item \textsc{Nearest}: Consider constraints $t_j - t_i$ in increasing order of the  number of intermediate events between $e_i$ and $e_j$: i.e,  $\left| \{ e_k\ |\ e_i \prec e_k \prec e_j \} \right| $.
\item \textsc{Distant}: Consider constraints $t_j - t_i$ in decreasing order of the number of intermediate events between $e_i$ and $e_j$.
\item \textsc{Random}: Consider constraints in a randomized order.
\item \textsc{Sound:}  The SOUND algorithm starts from the last node, checks for all time constraints $t_j - t_i \bowtie a$  that are dependent on $t_i$. If ALL of them are redundant, then we remove all the constraints, or else, if ANY of them are required, we keep all the constraints, because the clock at node $i$ is going to be required anyways.
\end{enumerate}

The intuition behind these heuristics is that the algorithm terminates faster depending on the problem and data. If most of the time constraints are locally dependent, then NEAREST should be chosen. If not, DISTANT should be chosen. RANDOM was included as a reference.

\begin{figure}[t]
    \centering
    \begin{minipage}[b]{0.45\linewidth}
    \centering
    \scalebox{.75}{

\begin{tikzpicture}[
    node distance=0.5\linewidth,
    every edge/.append style={thick, ->, >=stealth'},
    every node/.style={draw, rectangle, rounded corners, scale=0.8, font=\small, align=center}]
    ]
    \node(n1)[label={[align=center, xshift=-3mm]$0.0 \leq c_0 \leq 1.0$ \\ $c_1\colon=0$}]{$e_1$}; 
    \node(n2)[above right of=n1, label={[align=center, xshift=-5mm]$5.0 \leq c_1 \leq 15.0$ \\ $c_0\colon=0$}]{$e_2$};
    \node(n3)[right of=n2, label={[align=center, xshift=2mm]$15.0 \leq c_1 \leq 24.9$ \\ $c_2\colon=0$}]{$e_3$}; 
    \node(n4) at ($(n2)!0.5!(n3)$) [yshift=-20mm, label={[align=center]$0.0 \leq c_1 \leq 20.0$}]{$e_4$};
    \node(n5)[below right of=n3, label={[align=center, xshift=-4mm, yshift=1mm]$10.0 \leq c_2 \leq 11.0$ \\ $10.6 \leq c_0 \leq 30.3$}]{$e_5$}; 
    
    \path[->, line width=2pt] (n1) edge (n2)
    (n2) edge (n3)
    (n3) edge (n5)
    (n4) edge (n5)
    (n3) edge (n5)
    (n1) edge (n4);
\end{tikzpicture}}
    \subcaption{\textsc{nearest}}
    \label{fig:HeuristicNearest}
    \end{minipage}
    \hfill
    \begin{minipage}[b]{0.45\linewidth}
    \centering
    \scalebox{.75}{




\begin{tikzpicture}[
    node distance=0.5\linewidth,
    every edge/.append style={thick, ->, >=stealth'},
    every node/.style={draw, rectangle, rounded corners, scale=0.8, font=\small, align=center}]
    ]
    \node(n1)[label={[align=center, xshift=-3mm]$0.0 \leq c_0 \leq 1.0$ \\ $c_1\colon=0$}]{$e_1$}; 
    \node(n2)[above right of=n1, label={[align=center, xshift=-5mm]$5.0 \leq c_1 \leq 15.0$}]{$e_2$};
    \node(n3)[right of=n2, label={[align=center, xshift=0mm]$15.0 \leq c_1 \leq 24.9$ \\ $c_0\colon=0$}]{$e_3$}; 
    \node(n4) at ($(n2)!0.5!(n3)$) [yshift=-20mm, label={[align=center]$0.0 \leq c_1 \leq 20.0$}]{$e_4$};
    \node(n5)[below right of=n3, label={[align=center, xshift=-3mm, yshift=2mm]$10.0 \leq c_0 \leq 11.0$}]{$e_5$}; 
    
    \path[->, line width=2pt] (n1) edge (n2)
    (n2) edge (n3)
    (n3) edge (n5)
    (n4) edge (n5)
    (n3) edge (n5)
    (n1) edge (n4);
\end{tikzpicture}}
    \subcaption{\textsc{sound}}
    \label{fig:HeuristicBackwards}
    \end{minipage}
    \caption{Mined {\tpo}s using the \textsc{nearest} and \textsc{sound} heuristics.
    }
    \label{fig:RunningExampleEstimatedGraph}
\end{figure}
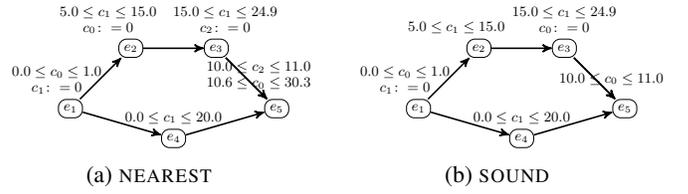

\begin{example}
Consider \Cref{ex:windshield-example} with the  new time constraints $t_1 \in [0, 1]$, $t_2 - t_1 \in [5, 15]$, $t_3-t_1 \in [15, 25]$, $t_4-t_1 \in [0, 20]$, and $t_5-t_3 \in [10, 11]$.
We randomly sampled 1000 timed traces that satisfy these constraints and  used the procedure described in this section to construct {\tpo}s, as shown in \Cref{fig:RunningExampleEstimatedGraph}. The TPO to the left uses the  \textsc{nearest} heuristic for redundancy elimination requiring three clocks, whereas the TPO to the  right uses the \textsc{sound} heuristic using just two clocks to explain the same data.
\end{example}

\subsubsection{Time Complexity}
Let $n=|\Pi|$. The algorithm solves a linear program (LP)  at each step whose time complexity is bounded by a polynomial over $n$. In the worst case, there are $O(n^2)$ pairs of time constraints (LP problems to solve).

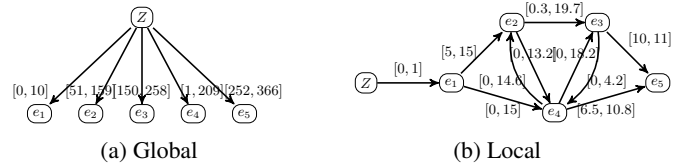
\begin{figure}[t]
    \centering
    \begin{minipage}[b]{0.45\linewidth}
        \centering
        \scalebox{.75}{

\begin{tikzpicture}[
    node distance=0.3\linewidth,
    every edge/.append style={thick, ->, >=stealth'},
    every node/.style={draw, rectangle, rounded corners, scale=0.8, font=\small, align=center}
    ]
    \node(Z){$Z$}; 
    \node(e3)[below of=Z, yshift=-10 mm, label={$[150, 258]$}]{$e_3$};
    \node(e2)[left of=e3, label={$[51, 159]$}]{$e_2$}; 
    \node(e1)[left of=e2, label={[xshift=-2mm]$[0, 10]$}]{$e_1$};
    \node(e4)[right of=e3, label={[xshift=2mm]$[1, 209]$}]{$e_4$};
    \node(e5)[right of=e4, label={[xshift=2mm]$[252, 366]$}]{$e_5$}; 
    
    \path[->, line width=2pt] 
    (Z) edge (e1)
    (Z) edge (e2)
    (Z) edge (e3)
    (Z) edge (e4)
    (Z) edge (e5);
\end{tikzpicture}}
        \subcaption{Global}
        \label{fig:cstnud_global}
    \end{minipage}
    \hfill
    \begin{minipage}[b]{0.45\linewidth}
        \centering
        \scalebox{.75}{

\begin{tikzpicture}[
    node distance=0.5\linewidth,
    every edge/.append style={thick, ->, >=stealth'},
    every node/.style={rectangle, rounded corners, scale=0.8, font=\small, align=center}]
    ]
    \node(Z)[draw]{$Z$};
    \node(n1)[draw, right of=Z, label={[align=center, xshift=-3mm]}]{$e_1$}; 
    \node(n2)[draw, above right of=n1, label={[align=center, xshift=-5mm]}]{$e_2$};
    \node(n3)[draw, right of=n2, label={[align=center, xshift=2mm]}]{$e_3$}; 
    \node(n4) at ($(n2)!0.5!(n3)$) [draw, yshift=-20mm, label={[align=center]}]{$e_4$};
    \node(n5)[draw, below right of=n3, label={[align=center, xshift=-4mm, yshift=1mm]}]{$e_5$}; 
    
    \path[->, line width=2pt] 
    (Z) edge node[above] {$[0, 1]$} (n1) 
    (n1) edge node[xshift=-5mm] {$[5, 15]$} (n2)
    (n2) edge node[above] {$[0.3, 19.7]$} (n3)
    (n2) edge node[above] {$[0, 13.2]$} (n4)
    (n4) edge[bend left] node[xshift=-3mm, yshift=-2mm] {$[0, 14.6]$} (n2)
    (n3) edge node[above, xshift=5mm] {$[10, 11]$} (n5)
    (n3) edge[bend left] node[xshift=3mm, yshift=-2mm] {$[0, 4.2]$} (n4)
    (n4) edge node[above] {$[0, 18.2]$} (n3)
    (n4) edge node[below] {$[6.5, 10.8]$} (n5)
    (n1) edge node[below] {$[0, 15]$} (n4);
\end{tikzpicture}}
        \subcaption{Local}
        \label{fig:cstnud_local}
    \end{minipage}

    \caption{Outputs by the CSTNUD mining algorithm (\url{https://github.com/matteozavatteri/cstnud-miner}) assuming (Left) all the events depend on the global clock $Z=0$ and (Right) each event depends on the previous event.}
    \label{fig:cstnud}
\end{figure}

\section{Experiments}

In this section, we evaluate our approach for mining {\tpo}s from timed trace data.
First, we compare our method against the most closest work, the CSTNUD mining algorithm.
Next, we run an ablative study on a benchmark to evaluate the clock allocation performance against a few heuristics. Subsequently, we show interesting results on two  datasets inspired from real-life applications: aircraft turnaround and Overcooked game.

\paragraph{Comparison against a CSTNUD mining algorithm}
As a comparison to \Cref{ex:windshield-example}, we ran the algorithm proposed by \cite{sciavicco2021mining}. Their method requires human annotations on event relationships, so we prepared two different result that can easily be derived from the raw data: 1) all events depending on the global clock (called "global") and each event depending on the previous event (called "local") and showed the resulting graphs in \Cref{fig:cstnud}. The globally-dependent graph entirely depends on the global clock and cannot mine the relationships between events. As a result, the timings constraints (durations) between events tend to become large. Whereas the locally-dependent graph is shown to mine the event relationships well but with unnecessary edges such as $e_2$ -- $e_4$ and $e_3$ -- $e_4$. In other words, they cannot handle parallelized tasks like ($e_2$, $e_3$) pair and $e_4$. Moreover, the CTSNDU mining algorithm can only mine relationships between neighboring edges that result in a one-clock model whereas our model can be viewed as a generalized version of their algorithm in which it can mine multiple-clock graphs.

\paragraph{Analysis on Synthetic TPO Data:}
In this experiment, we seek to understand (a) how the  running time of our procedure scales with increasing number of events/traces; and (b) the number of clocks generated by the various redundancy elimination heuristics.
We generated a bunch of random TPOs according to a method described in the appendix. The TPOs vary according to the number of events $n$. For each TPO, we randomly sampled 1000 traces.

\begin{figure}[t]
    \centering
    \begin{minipage}[b]{\linewidth}
        \centering
        \includegraphics[width=\linewidth]{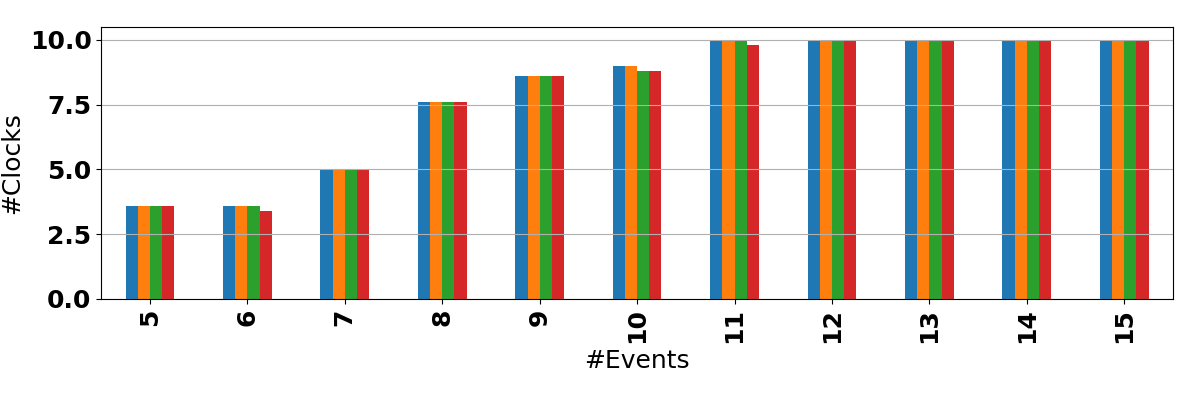}
        \subcaption{Number of Clocks}
        \label{fig:ClockAnalysis}
    \end{minipage}
    \begin{minipage}[b]{\linewidth}
        \centering
        \includegraphics[width=\linewidth]{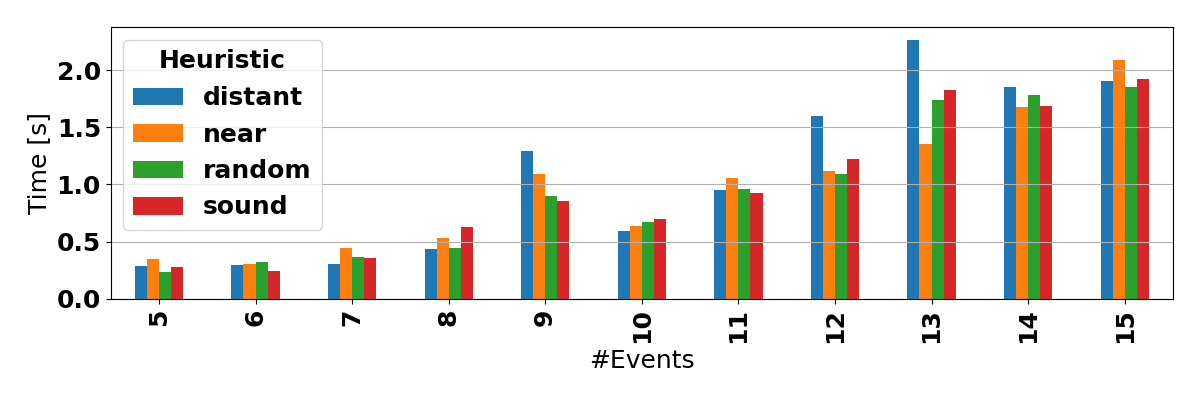}
        \subcaption{Computation Time}
        \label{fig:TimeComplexityAnalysis}
    \end{minipage}
    \caption{Benchmark results of the heuristic algorithms}
    \label{fig:AblativeStudy}
\end{figure}

\Cref{fig:ClockAnalysis} shows the average number of clocks identified over ten runs of this procedure. The number of clocks increases as the number of events increases. The heuristics yield similar number of clocks but the \textsc{sound} heuristic yields  fewer clocks (often one fewer clock).
\Cref{fig:TimeComplexityAnalysis} shows that the computation time increases with the number of events, as expected. The \textsc{sound} heuristic has a larger computation time due to the fact that some redundant constraints are retained.
The computation time varies on how many times the LP optimizations are called and how fast they find the solution. The more eliminations happen at the earlier stage, the fewer constraints remain in later LP. Often, however, the extra cost of upfront elimination does not provide enough of a payoff in the later stages. This is very much dependent on the nature of the data and constraints.

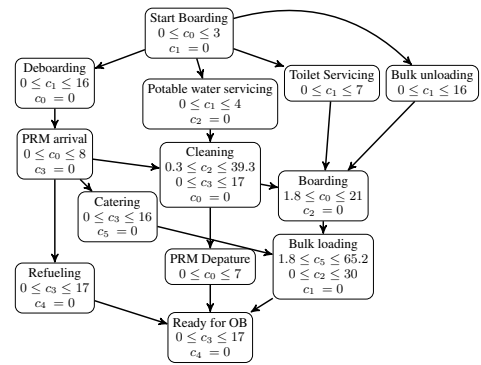
\begin{figure}[t]
    \centering
    \begin{minipage}[b]{\linewidth}
    \centering
    \scalebox{.7}{

\begin{tikzpicture}[
    node distance=0.2\linewidth,
    every edge/.append style={thick, ->, >=stealth'},
    every node/.style={draw, rectangle, rounded corners, scale=0.8, font=\small, align=center}]

  \node  (e1)[]{Start Boarding\\$0\leq c_0 \leq 3$\\$c_1\:=0$};
  \node  (e2)[below left of=e1, xshift=-2cm]{Deboarding\\$0\leq c_1 \leq 16$\\$c_0\:=0$};
  \node  (e3)[below of=e1, xshift=0.5cm]{Potable water servicing\\$0\leq c_1 \leq 4$\\$c_2\:=0$};
  \node  (e4)[below right of=e1, xshift=2.2cm]{Toilet Servicing\\$0\leq c_1 \leq 7$};
  \node  (e5)[right of=e4, xshift=0.7cm]{Bulk unloading\\$0\leq c_1 \leq 16$};
  \node  (e6)[below of=e2]{PRM arrival\\$0\leq c_0 \leq 8$\\$c_3\:=0$};
  \node  (e7)[below right of=e6, xshift=0.3cm, yshift=-0.3cm]{Catering\\$0\leq c_3 \leq 16$\\$c_5\:=0$};
  \node  (e8)[below of=e3]{Cleaning\\$0.3\leq c_2 \leq 39.3$\\$0\leq c_3 \leq 17$\\$c_0\:=0$};
  \node  (e9)[right of=e8, xshift=1cm, yshift=-0.5cm]{Boarding\\$1.8\leq c_0 \leq 21$\\$c_2\:=0$};
  \node  (e10)[below of=e8, yshift=-0.5cm]{PRM Depature\\$0\leq c_0 \leq 7$};
  \node  (e11)[below of=e9]{Bulk loading\\$1.8\leq c_5 \leq 65.2$\\$0\leq c_2\leq 30$\\$c_1\:=0$};
  \node  (e12)[below of=e6, yshift=-1.5cm]{Refueling\\$0\leq c_3 \leq 17$\\$c_4\:=0$};
  \node  (e13)[below of=e10]{Ready for OB\\$0\leq c_3 \leq 17$\\$c_4\:=0$};
  
  \draw
  (e1) edge (e2)
  (e1) edge (e3)
  (e1) edge (e4)
  (e1) edge[bend left] (e5)
  (e2) edge (e6)
  (e6) edge (e7)
  (e6) edge (e8)
  (e3) edge (e8)
  (e4) edge (e9)
  (e5) edge (e9)
  (e8) edge (e9)
  (e8) edge (e10)
  (e7) edge (e11)
  (e9) edge (e11)
  (e6) edge (e12)
  (e10) edge (e13)
  (e11) edge (e13)
  (e12) edge (e13)
  ;
\end{tikzpicture}}
    \end{minipage}
\caption{Mined \tpo for aircraft turnaround.}
\label{fig:AircraftTrunAround}
\end{figure}


\paragraph{Aircraft Turnaround Example:}
Next, we evaluate our algorithm on the processes involved in the turnaround of an aircraft at a gate. Aircraft turnaround is a critically important process that affects the operating costs of airlines. It involves a series of tasks such as deboarding, cleaning, refueling and boarding with happens-before orders. For instance, cleaning must be performed after deboarding. However, refueling can be performed in parallel with cleaning.
We defined a \tpo of the aircraft turnaround operations using data synthesized from the information presented in \citet{nosedal2018causal}. In particular, we use \textit{average time that led to delays} as the maximum time for each operation. Timestamps were sampled from truncated normal distributions, as specified by \citet{nosedal2018causal}.
We synthesized $1000$ timed traces and evaluated our algorithm against the SMT-based timed automata inference algorithm \cite{tappler2022timed}.
The SMT-based algorithm did not terminate over the original data set (timed out due to expensive calls to SMT solvers). We had to reduce the data size to just 20 timed traces in order to get the algorithm to run. However, the  result fails to capture the timing constraints present in the original problem. The  RTI+ algorithm \cite{verwer2012efficiently} also fails to run on the reduced data.

In contrast, the \tpo mined using our algorithm shown in \Cref{fig:AircraftTrunAround} respects all original precedence orders, along with the maximum time duration for each operation. Furthermore, notice that the algorithm identified new precedence orders such as ``Toilet servicing" happens-before ``Boarding" due to the relation between the two time constraints. The TPO just requires $6$ clocks in all. This example shows that our algorithm can be applied to realistic scenarios.

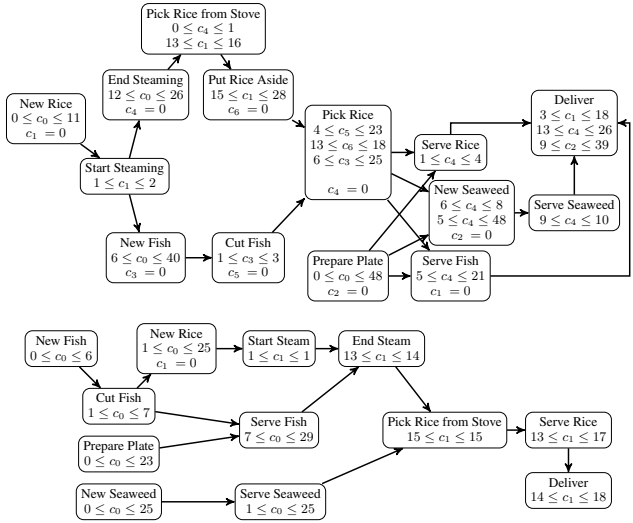
\begin{figure}[t]
    \begin{tabular}{c}
    \begin{minipage}[b]{\linewidth}
        \centering
        \scalebox{.7}{

\begin{tikzpicture}[
    node distance=0.23\linewidth,
    every edge/.append style={thick, ->, >=stealth'},
    every node/.style={draw, rectangle, rounded corners, scale=0.8, font=\small, align=center}]

  \node  (e1)[]{New Rice\\$0\leq c_0 \leq 11$\\$c_1\:=0$};
  \node  (e2)[below right of=e1, xshift=0.5cm]{Start Steaming\\$1\leq c_1 \leq 2$};
  \node  (e3)[above of=e2, xshift=0.5cm]{End Steaming\\$12\leq c_0 \leq 26$\\$c_4\:=0$};
  \node  (e4)[below of=e2, xshift=0.5cm]{New Fish\\$6\leq c_0 \leq 40$\\$c_3\:=0$};
  \node  (e5)[above right of=e3, yshift=0.2cm]{Pick Rice from Stove\\$0\leq c_4 \leq 1$\\$13\leq c_1\leq 16$};
  \node  (e6)[right of=e3, xshift=0.5cm]{Put Rice Aside\\$15\leq c_1 \leq 28$\\$c_6\:=0$};
  \node  (e7)[right of=e4, xshift=0.5cm]{Cut Fish\\$1\leq c_3 \leq 3$\\$c_5\:=0$};
  \node  (e8)[below right of=e6, xshift=1cm]{Pick Rice\\$4\leq c_5 \leq 23$\\$13\leq c_6 \leq 18$\\$6\leq c_3 \leq 25$\\\\$c_4\:=0$};
  \node  (e9)[below of=e8, yshift=-1cm]{Prepare Plate\\$0\leq c_0 \leq 48$\\$c_2\:=0$};
  \node  (e10)[right of=e8, xshift=0.5cm]{Serve Rice\\$1\leq c_4 \leq 4$};
  \node  (e11)[below of=e10, xshift=0.5cm, yshift=0.5cm]{New Seaweed\\$6\leq c_4 \leq 8$\\$5\leq c_4\leq 48$\\$c_2\:=0$}; 
  \node  (e12)[right of=e9, xshift=0.5cm]{Serve Fish\\$5\leq c_4 \leq 21$\\$c_1\:=0$};
  \node  (e13)[right of=e11, xshift=0.5cm]{Serve Seaweed\\$9\leq c_4 \leq 10$};
  \node  (e14)[above of=e13, yshift=0.2cm]{Deliver\\$3\leq c_1 \leq 18$\\$13\leq c_4 \leq 26$\\$9\leq c_2 \leq 39$};
  
  \draw
  (e1) edge (e2)
  (e2) edge (e3)
  (e2) edge (e4)
  (e3) edge (e5)
  (e4) edge (e7)
  (e5) edge (e6)
  (e6) edge (e8)
  (e7) edge (e8)
  (e8) edge (e10)
  (e8) edge (e11)
  (e8) edge (e12)
  (e9) edge (e10)
  (e9) edge (e11)
  (e9) edge (e12)
  (e11) edge (e13)
  (e13) edge (e14)
  ;
  \draw[thick, ->, >=stealth'] (e10) |- (e14);
  \draw[thick, ->, >=stealth'] (e12) -- ([shift={(3.4,0)}]e12.center) |- (e14);
  
\end{tikzpicture}}
    \end{minipage}\\[5pt]
    \begin{minipage}[b]{\linewidth}
        \centering
        \scalebox{.7}{

\begin{tikzpicture}[
    node distance=0.23\linewidth,
    every edge/.append style={thick, ->, >=stealth'},
    every node/.style={draw, rectangle, rounded corners, scale=0.8, font=\small, align=center}]

  \node  (e1)[]{New Fish\\$0\leq c_0 \leq 6$};
  \node  (e2)[below right of=e1]{Cut Fish\\$1\leq c_0 \leq 7$};
  \node  (e3)[above right of=e2]{New Rice\\$1\leq c_0 \leq 25$\\$c_1\:=0$};
  \node  (e4)[right of=e3, xshift=0.5cm]{Start Steam\\$1\leq c_1 \leq 1$};
  \node  (e5)[below of=e2, yshift=0.8cm]{Prepare Plate\\$0\leq c_0 \leq 23$};
  \node  (e6)[below of=e4]{Serve Fish\\$7\leq c_0 \leq 29$};
  \node  (e7)[below of=e5, yshift=0.8cm]{New Seaweed\\$0\leq c_0 \leq 25$};
  \node  (e8)[right of=e7, xshift=1.9cm]{Serve Seaweed\\$1\leq c_0 \leq 25$};
  \node  (e9)[right of=e4, xshift=0.5cm]{End Steam\\$13\leq c_1 \leq 14$};
  \node  (e10)[right of=e6, xshift=2cm]{Pick Rice from Stove\\$15\leq c_1 \leq 15$};
  \node  (e11)[right of=e10, xshift=1cm]{Serve Rice\\$13\leq c_1\leq 17$};
  \node  (e12)[below of=e11, yshift=0.5cm]{Deliver\\$14\leq c_1 \leq 18$};
  
  \draw
  (e1) edge (e2)
  (e2) edge (e3)
  (e2) edge (e6)
  (e3) edge (e4)
  (e5) edge (e6)
  (e7) edge (e8)
  (e4) edge (e9)
  (e6) edge (e9)
  (e8) edge (e10)
  (e9) edge (e10)
  (e10) edge (e11)
  (e11) edge (e12)
  ;
\end{tikzpicture}}
    \end{minipage}
    \end{tabular}
\caption{ Overcooked gameplay analysis: (Top) TPO specification for a beginner player, (Bottom) TPO specification of a professional player.}\label{fig:overcooked-tpo}
\end{figure}

\paragraph{Overcooked Example:} Overcooked is a multiplayer game that simulates a busy restaurant kitchen, requiring
players to collaborate on producing numerous
plates of food according to a fixed recipe against
timing constraints.
We tested our algorithm on publicly available game play videos on YouTube.
We analyzed the difference between the
beginner\footnote{\url{https://www.youtube.com/watch?v=jTrenjjZDtA&t=668s}} versus professional\footnote{\url{https://www.youtube.com/watch?v=YcnpWo4Y01M&t=60s}} gamers on Overcooked 2.
The task of the game is to repeatedly make Sushi plates  from seaweed, sliced fish, and cooked rice, serve them  to customers in a target area.
There is no serving order between the three ingredients, but there are strict orders on how each ingredient is prepared.
We manually annotated the video with event labels describing the actions of the players.
We were able to identify eight timed traces in the beginner's play and  collected the same number of traces from the professional's play. The mined {\tpo}s are shown in \Cref{fig:overcooked-tpo}.

In both cases, strict orders are correctly identified. For example, a fish must be cut before serving on a plate.
The difference between the two is the order of the parallelizable tasks.
Interestingly, the beginners start with \textit{cooking rice} and then \textit{cutting fish}, whereas the professionals start with \textit{cutting fish} and then \textit{cooking rice}.
This is because beginners prepare dish one by one, and hence they must start with rice, which takes the longest time to prepare, whereas the professionals cook in a batch and the fish comes first in this strategy.

In terms of time constraints, strict constraints are imposed, such as (1) fish/seaweed must be cut/served immediately after a new object is taken out of the box (2) rice must be steamed for about 13-16 seconds in both figures. In contrast, a new object (plate, rice, and fish) can be taken out of the box at any time in the scene (with a large bound in $c_0$).
Furthermore, a time constraint between serving fish and rice (Serve Rice $4 \leq c_1 \leq 14$) in \Cref{fig:overcooked-tpo} (bottom) specifies that parallel tasks must take similar times, so the dish can be delivered immediately after (Deliver $1 \leq c_4 \leq 2$).
This experiment shows that our \tpo mining algorithm provides an interpretable representation for a task solely based on a small amount of  data.

\section{Discussion}

\subsubsection{Pipeline Workflow: Repetitive Events}
In our problem setting, we assumed that the log data contains a neatly classified set of traces and each trace contains only a unique set of events. However, in reality, a log consists of a set of mixed traces sorted by timestamps and there could be multiple occurrences of the same event in a trace. For example, an automotive assembly line manufactures multiple cars concurrently and the same events (e.g., install a door) appear multiple times in each trace. 
To split the log into a set of traces, we need to identify the counts of each event appearing in a trace $x \in \mathbb{Z}^n_{>0}$ and split the log accordingly. To do so, we formulate the problem as an integer programming. Given a number of products being manufactured $k$ and a log, minimize $x$, such that, $k \cdot I \cdot x  + I \cdot y = b $, $x>0, y>=0$ and $y<k$, 
where $y\in \mathbb{Z}^n_{\geq0}$ is a vector of variables representing the remaining events in a trace and $b\in \mathbb{Z}^n_{>0}$ is the counts of events in the log. Then, we greedily split the log from the top with each trace containing $x$ counts of events.

\subsubsection{Loops}
In our formulation, we cannot model loops in the partial order graph. However, in reality, a log can come from a workflow that requires certain events to loop. For, example, cracking three eggs can be represented as repetitions of 
\emph{NewEgg} and \emph{Crack} events. To learn a TPO from such data, the naive way is to relabel the repetitive events with unique events, e.g., \emph{NewEgg1}, \emph{NewEgg2}. 
Once the partial among other events are identified, then the repeated events can be folded. For example, \emph{Crack1} can be relabeled back too \emph{Crack} and add an edge to \emph{NewEgg} to form a loop. We will detail this idea in an extended version of this paper.
\section{Conclusions}

Thus, we have introduced TPOs that integrate the partial orders with ideas from timed automata to express the task orders and timing constraints between events. 
We have analyzed the expressivity of {\tpo}s leading to a procedure for mining TPOs from data. Experiments demonstrate how mining TPOs from process data can yield  useful insights for important workflows inspired by real-life manufacturing processes.
\newpage
\appendix
\section{Proofs of Theorems}

\subsection{Proof of Theorem~\ref{Theorem:tpo-expressivity-race-free}}

\begin{proof}(Sketch) We will construct $\varphi$ starting from the initial assertion $\varphi_0: \psi_{POR}$ and consider each event $e_i$ in turn for $i=1,\ldots, n$. It is easy to see that $\psi_{POR}$ itself is an assertion of the form required by the statement of the theorem.
Let $\varphi_{i-1}$ be the assertion after events $e_1, \ldots, e_{i-1}$ have been considered. We will add conjuncts to $\varphi_{i-1}$ one for each conjunct in the guard $G(e_i)$ for event $e_i$ to yield $\varphi_i$.

Consider each conjunct $c_j \bowtie l_j$ of the guard $G(e_i)$ (the guard of event $e_i$), wherein $\bowtie \in\{ \leq, \geq \}$. Consider the set of all events that reset $c_j$ and precede $e_i$ in the partial order: $E_{i,j} = \{ e_k\ |\ e_k\ \text{resets the clock}\ c_j\ \land\ e_k \prec e_i \}$. Consider any timed trace such that 
events happen at times $t_1, \ldots, t_n$. When event $e_i$ happens, one of two cases may apply: (a) the clock $c_j$ has never been reset; or (b) the clock $c_j$ was reset by some preceding event $e_k$. Since the TPO is race-free, we note that for the case (a) $E_{i,j} = \emptyset$ and for case (b) $e_k$ is the maximal element in $E_{i,j}$ with respect to $\prec$. 

If $E_{i,j}$ is empty, we infer the constraint $t_i \bowtie l_j$ to the assertion $\varphi_{i-1}$: the timing of event $e_i$ must satisfy the clock guard and in this case, the clock has not been reset since the start of the process. On the other hand, if $E_{i,j}$ is non-empty, the clock $c_j$ was last reset at time $t_k$ corresponding to event $e_k$ that is maximal in $E_{i,j}$ w.r.t the $\prec$ order. We add the constraint $t_i - t_k \bowtie l_j$ to represent the guard. Thus we obtain $\varphi_i$ after considering each conjunct of $G(e_i)$. Note that the conjuncts added all have the form $t_i \bowtie l_j$ or $t_i - t_k \bowtie l_j$. The overall constraint $\varphi = \varphi_n$  has the form specified in the theorem and represents all possible timed traces that are compatible with the TPO. 
\end{proof}

\section{Experiment Setup}
Our algorithm was implemented using Python 3.10 on a MacBook Pro (16 GB 2133 MHz, 2.3 GHz Dual Core Intel Core i5). 

\section{Random Generation of {\tpo}s}
We used a verification tool for Real Time Systems called \textsc{UPPAAL}\footnote{\url{https://uppaal.org/}} to randomly sample timed traces from timed-automata. To do so, we translated {\tpo}s to compatible timed-automata.
To generate a random \tpo, we randomly generated a Directed Acyclic Graph (DAG) for and performed a transitive reduction to obtain a transitive reduced \po. The number of events can vary from 1 to $n$ depending on this process. 
Subsequently, we translated the graph to an automaton that accommodates all linearizations of the \po and randomly generated time bounds using three clocks such that there always exist a feasible path from any node to the final node without any deadlocks. 
To always guarantee that such a path exists, we construct feasible \text{zones} for all nodes and edges as we generate new time bounds. A zone is defined as the bounded area defined by all clocks and the differences between all pairs of clocks in the timed automata, i.e., 
$\big( \bigwedge_{c_i\in C} c_i \bowtie a_i \big)$ $\wedge$ $\big( \bigwedge_{c_i, c_j \in C} c_i - c_j \bowtie a_{i,j} \big)$. 
The inequalities form a polyhedron where all possible time stamps for an event lies in this zone \cite{gastin2018reachability, bouyer2022zone}. To ensure feasible paths, we generate a new lower bound that lies between the precedent zone bounds and a new upper bound that is greater than or equal to a precedent upper bound. It gets tricky when multiple incoming edges intersect at a node. Each edge has its own zone, and we must take a union of the zones to guarantee a feasible path from all incoming edges. To do so, we take the minimum and maximum of the clock differences $c_i - c_j \bowtie a_{i,j}$ and update all clock bounds $c_i \bowtie a_i$ accordingly. This results in a feasible timed automaton with arbitrary number of nodes that is smaller than $n^2$. 
We used the seed of 1337 and increased by one as we increased the number of traces.

\begin{table}[t]
    \centering
    \begin{tabular}{ll}
        \hline
        Name & Value \\ \hline
        nrTrainingTraces & 20 \\
        maxNrLocations & 15 \\
        maxNrEdges & 30 \\
        edgesPerLocation & 5 \\
        kUrgency & 30 \\
        maxGuardConstant & 30 \\
        incremental & true \\
        bfsMode & true \\
        discreteFirst & true \\
        solver & SMTINTERPOL \\
        theory & Real \\
        roundingFactor & 1 \\
        skipQualityEval & true \\ \hline
    \end{tabular}
    \caption{Hyperparameters for the SMT-based method}
    \label{tab:my_label}
\end{table}
\begin{table}[t]
    \centering
    \begin{tabular}{ll}
        \hline
        Name & Value \\ \hline
        heuristic-name & rtiplus \\
        data-name & rtiplus\_data \\
        state\_count & 0 \\
        symbol\_count & 0\\
        satdfabound & 2000 \\
        largestblue & 1 \\
        sinkson & 0 \\
        sinkcount & 3 \\
        confidence\_bound & 0.1 \\
        extend & 0 \\
        finalred & 0 \\
        finalprob & 0 \\ \hline
    \end{tabular}
    \caption{Hyperparameters for the RTI+ algorithm}
    \label{tab:rtiplus_hyperparameters}
\end{table}

\section{Comparative Methods}
We ran the SMT-based method\footnote{\url{https://github.com/mtappler/smt-ta-learning}} that is implemented by the original author, with the following hyperparameters.

We compared against the RTI+ algorithm implemented in the \textsc{FlexFringe} library\footnote{\url{https://github.com/tudelft-cda-lab/FlexFringe}} that is also implemented by the original author. We ran the RTI+ algorithm with the following hyperparameters.

\newpage

\bibliography{references}

\end{document}